\documentclass[prd,reprint,nofootinbib,preprintnumbers,showpacs]{revtex4-1}
\usepackage{CJK}
\usepackage{hyperref,amsmath,amssymb,graphicx,bm,bbm,multirow}
\newcommand{\arXivid}[1]{\href{http://arxiv.org/abs/#1}{arXiv:#1}}

\begin{document}
\preprint{PUPT-2426}
\title{Analytic quantum critical points from holography}
\author{Jie Ren}
\affiliation{Department of Physics, Princeton University, Princeton, NJ 08544}
\date{\today}
\begin{abstract}
  We find exact, analytic solutions of the Klein-Gordon equation for a scalar field in the background of the extremal Reissner-Nordstr\"{o}m-AdS$_5$ black hole. The Green's function near a quantum critical point for a strongly coupled system can be extracted holographically from an exact solution for the scalar at zero frequency ($\omega$), but arbitrary momentum ($\mathbf{k}$), mass, and charge. By examining the Green's function near $\omega=0$, there are two types of instability: the first one is triggered by a zero mode, and gives a hybridized critical point; the second one is triggered by the instability of the IR geometry, and gives a bifurcating critical point. The two types of instability can happen at the same time, and give a mixed critical point. Without tuning an extra parameter, only the second type of instability can happen at $\mathbf{k}=0$. At the critical point with the superfluid velocity, the scalar can develop either type of instability, depending on the parameters. The zero mode can also be obtained by tuning a double trace deformation. The phase diagrams can be analytically drawn.
\end{abstract}
\maketitle

\section{Introduction\label{sec:intro}}
Phase transitions for some strongly interacting systems can be modeled by the gauge/gravity duality \cite{Maldacena:1997re,Gubser:1998bc,Witten:1998qj}. For example, holographic superconductors have been constructed in terms of asymptotic anti-de Sitter (AdS) spacetimes \cite{Hartnoll:2008vx,Hartnoll:2008kx}. In this paper, we consider quantum phase transitions, i.e., the phase transitions that happen at zero temperature. Understanding the quantum critical points is a significant challenge in condensed matter physics, such as non-conventional superconductors. In the gravity description, a phase transition happens when a scalar field in an AdS background develops an instability. Solving the Klein-Gordon equation for the scalar field in the bulk gives the Green's function of a scalar operator in the boundary. The critical point can be identified by the non-analyticity of the Green's function at $\omega=0$.

Previous studies have made major conceptual progresses in understanding the quantum critical points in terms of the Reissner-Nordstr\"{o}m (RN) black hole in AdS$_4$ \cite{Iqbal:2011aj} (for another system, see Ref.~\cite{Evans:2010np}; the low-energy effective field theory is also identified in Ref.~\cite{Jensen:2011af}). The extremal RN-AdS black hole has an AdS$_2$ factor in its near horizon (hereafter IR for infrared) geometry \cite{Faulkner:2009wj}. This AdS$_2$ factor plays an essential role in the properties of the system, and defines a universal intermediate energy phase \cite{Iqbal:2011in,Faulkner:2010tq}. The behavior of the system near the quantum critical point is encoded in the Green's function near $\omega=0$, which contains both UV and IR data. The UV data can be solved from the Klein-Gordon equation at $\omega=0$ in the bulk, and the IR data can be analytically solved from the Klein-Gordon equation at arbitrary $\omega$ in AdS$_2$ \cite{Iqbal:2011aj,Faulkner:2009wj}. The zero mode, which triggers the onset of the instability of the system, belongs to the UV data, and relies on numerical calculations in AdS$_4$.

We find that if we use AdS$_5$ instead, the Klein-Gordon equation at $\omega=0$ can be analytically solved; the Green's function captures essential features of the RN-AdS system. We consider the standard/alternative quantization first. For a scalar field with mass $m$ and charge $q$, the zero modes are solved as
\begin{equation}
\nu_k=\frac{q}{\sqrt{3}}-n_\pm-\frac{\Delta_\pm-1}{2},
\end{equation}
where $n_\pm$ is a nonnegative integer, $\Delta_\pm$ is the scaling dimension of the boundary operator, and $\nu_k$ is the IR scaling exponent:
\begin{align}
\Delta_\pm &:=2\pm\sqrt{m^2+4},\\
\nu_k &:=\frac{1}{2\sqrt{3}}\sqrt{m^2+k^2-2q^2+3}.
\end{align}
The zero modes are always at nonzero $\mathbf{k}$, and the $n_\pm=0$ mode triggers the onset of the instability. The instabilities and the corresponding quantum critical points are classified as follows (the name and interpretation of the quantum critical points are from Ref.~\cite{Iqbal:2011aj}):
\begin{itemize}
  \item The first type of instability is triggered by a zero mode, which exists only if $q$ is large enough. This instability gives a hybridized critical point, which is described by an order parameter in a Ginzburg-Landau sector hybridized with a strongly coupled sector, the CFT$_1$ dual to the IR AdS$_2$.
  \item The second type of instability happens when the IR scaling exponent $\nu_k$ becomes imaginary, which implies the instability of the IR geometry. This instability gives a bifurcating critical point, for which the Green's function bifurcates into the complex plane.
  \item The two types of instability can happen at the same time, and give a mixed critical point, such as a marginal critical point, which is described by a marginal term.
\end{itemize}
We then study the quantum critical points at $\omega=0$ and $\mathbf{k}=0$. Without introducing an extra parameter, we can only have a bifurcating critical point. We can tune the superfluid velocity to reach all three quantum critical points. The phase diagram can be analytically obtained for the onset of the instability from the normal phase.

Instead of the superfluid velocity, the zero mode can also be obtained by tuning another parameter $\kappa_+$, the coefficient of a double trace deformation in the CFT, giving a hybridized critical point \cite{Iqbal:2011aj,Faulkner:2010gj}. The analytic result allows us to draw the phase diagram for arbitrarily large $m^2$, where the numerical result is difficult to achieve. The Green's function (or susceptibility) for various critical points can be obtained.

This paper is organized as follows. In Sec.~\ref{sec:sol}, we give the general solution of the Klein-Gordon equation at $\omega=0$, and the Green's function near $\omega=0$. In Sec.~\ref{sec:xi}, we classify the instabilities according to the parameters $m^2$ and $q$, and give the critical values of the superfluid velocity. In Sec.~\ref{sec:kappa}, we consider the parameter of the double trace deformation, and draw the phase diagram. In Sec.~\ref{sec:sum}, we conclude with some discussions.

\section{Solution for the Klein-Gordon equation\label{sec:sol}}
The extremal RN-AdS$_5$ black hole in Poincar\'{e} coordinates is\footnote{The action is $S=\int d^5x\sqrt{-g}\,(R-12-\frac{1}{4}F^2)$.}
\begin{gather}
ds^2=\frac{1}{z^2}\left(-f(z)dt^2+d\mathbf{x}^2+\frac{dz^2}{f(z)}\right),\\
f=1-3z^4+2z^6,\qquad A_t=\sqrt{6}\,(1-z^2),
\end{gather}
where $\mathbf{x}=(x_1,x_2,x_3)$, and the gauge potential is $A=A_tdt$. We set the AdS radius $L=1$, and the horizon is at $z_h=1$.

To obtain the Green's function for a scalar operator in the dual CFT, we will solve the Klein-Gordon equation for a scalar field $\Phi$. After the Fourier transform
\begin{equation}
\Phi(z,x^\mu)=\int\frac{d\omega d^3\mathbf{k}}{(2\pi)^4}e^{-i\omega t+i\mathbf{k}\cdot\mathbf{x}}\phi(z),
\end{equation}
the equation of motion for $\phi$ is
\begin{equation}
\phi''+\left(\frac{f'}{f}-\frac{3}{z}\right)\phi'+\left(\frac{(\omega+qA_t)^2}{f^2}-\frac{k^2}{f}
-\frac{m^2}{z^2f}\right)\phi=0,\label{eq:KG}
\end{equation}
where we assume $\mathbf{k}=(k,0,0)$ without loss of generality, $q>0$, and $m^2$ is above the Breitenlohner-Freedman (BF) bound \cite{Breitenlohner:1982jf}: $m^2\geq m_\text{BF}^2=-4$. The horizon $z=1$ is an irregular singularity for this equation. The in-falling boundary condition near the horizon $z=1$ is
\begin{equation}
\phi\sim W_{-\frac{iq}{\sqrt{6}},\nu_k}\Bigl(-\frac{i\omega}{6(1-z)}\Bigr),\label{eq:phiin}
\end{equation}
where $W_{\lambda,\mu}(z)$ is a Whittaker function. The asymptotic behavior near the AdS boundary is\footnote{When $\Delta_+-\Delta_-=2n$, where $n=1$, $2$, $\cdots$, there are extra terms $bz^{\Delta_+}\ln z\,(1+\cdots)$. When $\Delta_+=\Delta_-$, $\phi=Az^2\ln z+Bz^2+\cdots$.}
\begin{equation}
\phi=Az^{\Delta_-}(1+\cdots)+Bz^{\Delta_+}(1+\cdots).
\end{equation}
The retarded Green's function is\footnote{We use the same normalization as in Ref.~\cite{Iqbal:2011aj}. By another normalization, the Green's function is $G=(2\Delta_+-4)B/A$.}
\begin{equation}
G=\frac{B}{A}.
\end{equation}
When $-4\leq m^2\leq -3$, there is an alternative quantization, by which the Green's function is $G=A/B$ \cite{Klebanov:1999tb}.

\begin{figure}
  \centering
  \includegraphics[]{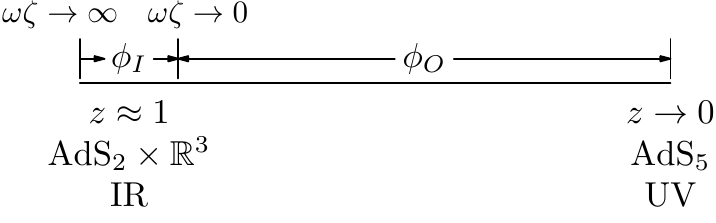}
  \caption{\label{fig:match} The inner (near horizon) and outer regions, where the solutions of the Klein-Gordon equation are denoted by $\phi_I$ and $\phi_O$, respectively.}
\end{figure}

To study the instability near a quantum critical point, we need to solve the Green's function near $\omega=0$. When it is sufficiently close to the extremal horizon, $\omega$-dependent terms cannot be treated as small perturbations no matter how small $\omega$ is. In Ref.~\cite{Faulkner:2009wj}, a systematic method is developed for treating the extremal black hole system. We divide the geometry into inner and outer regions, as shown in Fig.~\ref{fig:match}. The inner region refers to the IR (near horizon) geometry, in which the Klein-Gordon equation can be exactly solved as Eq.~\eqref{eq:phiin}. The outer region refers to the remaining geometry, in which we can make perturbations for small $\omega$. Then we need to match the inner and outer regions.

In the inner region, the IR Green's function $\mathcal{G}_k(\omega)$ is solved as Eq.~\eqref{eq:gir} (see Appendix~\ref{sec:gir}) \cite{Faulkner:2009wj}. In the outer region, the solution at small $\omega$ can be written as
\begin{equation}
\phi(z)=\eta_+(z)+\mathcal{G}_k(\omega)\eta_-(z),
\end{equation}
where
\begin{equation}
\eta_\pm=\eta_\pm^{(0)}+\omega\eta_\pm^{(1)}+\mathcal{O}(\omega^2).
\end{equation}
At the leading order, the asymptotic behavior near the horizon $z=1$ is
\begin{equation}
\eta_\pm^{(0)}\to[12(1-z)]^{-1/2\pm\nu_k},\label{eq:etah}
\end{equation}
which we also use to fix the normalization of the solution. The asymptotic behavior near the AdS boundary $z=0$ is
\begin{equation}
\eta_\pm^{(0)}\to a_\pm^{(0)}z^{\Delta_-}+b_\pm^{(0)}z^{\Delta_+}.\label{eq:etab}
\end{equation}
The Green's function to the first order in $\omega$ is \cite{Faulkner:2009wj}
\begin{equation}
G(\omega,k)=\frac{b_+^{(0)}+\omega b_+^{(1)}+{\cal G}_k(\omega)\bigl(b_-^{(0)}+\omega b_-^{(1)}\bigr)}
{a_+^{(0)}+\omega a_+^{(1)}+{\cal G}_k(\omega)\bigl(a_-^{(0)}+\omega a_-^{(1)}\bigr)}.\label{eq:green}
\end{equation}
Note that if we consider a neutral scalar, the first order terms in $\omega$ are zero, and then we need to expand to the second order. The analytic solution of $\phi$ at $\omega=0$ below gives the leading order of the Green's function. By perturbation around $\omega=0$, we can obtain the higher-order coefficients. The Green's function can be generalized to nonzero temperature when $T<<\mu$ (chemical potential) by replacing the IR Green's function $\mathcal{G}_k(\omega)$ with Eq.~\eqref{eq:girT} in Appendix~\ref{sec:gir}. The results we just described can be found in Ref.~\cite{Faulkner:2009wj}. We now go beyond Ref.~\cite{Faulkner:2009wj} and obtain analytic solutions for $a_\pm^{(0)}$ and $b_\pm^{(0)}$.

When $\omega=0$, we can solve $\phi$ in terms of hypergeometric equations. The general solution of $\phi$ for Eq.~\eqref{eq:KG} at $\omega=0$ is\footnote{When $\Delta_+$ is an integer, the two hypergeometric functions in Eq.~\eqref{eq:sol} are linearly dependent. We can choose another two linearly independent solutions as Eq.~\eqref{eq:sol2} in Appendix~\ref{sec:notes}.}
\begin{widetext}
\begin{align}
\phi(z)= &C_1z^{\Delta_-}\frac{(1-z^2)^{-1/2+\nu_k}}{(2z^2+1)^{-1/2+\nu_k+\Delta_-/2}}\,
{_2F_1}\Bigl(\frac{\Delta_--1}{2}+\nu_k-\frac{q}{\sqrt{3}},\,\frac{\Delta_--1}{2}+\nu_k+\frac{q}{\sqrt{3}};\,
\Delta_--1;\,\frac{3z^2}{2z^2+1}\Bigr)\nonumber\\
+ &C_2z^{\Delta_+}\frac{(1-z^2)^{-1/2+\nu_k}}{(2z^2+1)^{-1/2+\nu_k+\Delta_+/2}}\,
{_2F_1}\Bigl(\frac{\Delta_+-1}{2}+\nu_k-\frac{q}{\sqrt{3}},\,\frac{\Delta_+-1}{2}+\nu_k+\frac{q}{\sqrt{3}};\,
\Delta_+-1;\,\frac{3z^2}{2z^2+1}\Bigr).\label{eq:sol}
\end{align}
The asymptotic behavior near the horizon $z\to 1$ is
\begin{align}
\phi\to &\Bigl[\frac{2}{3}(1-z)\Bigr]^{-1/2+\nu_k}\left(\frac{C_1\,3^{-\Delta_-/2}\Gamma(\Delta_--1)\Gamma(-2\nu_k)}
{\Gamma\bigl(\frac{\Delta_--1}{2}-\nu_k+\frac{\sqrt{3}}{3}q\bigr)
\Gamma\bigl(\frac{\Delta_--1}{2}-\nu_k-\frac{q}{\sqrt{3}}\bigr)}
+\frac{C_2\,3^{-\Delta_+/2}\Gamma(\Delta_+-1)\Gamma(-2\nu_k)}
{\Gamma\bigl(\frac{\Delta_+-1}{2}-\nu_k+\frac{q}{\sqrt{3}}\bigr)
\Gamma\bigl(\frac{\Delta_+-1}{2}-\nu_k-\frac{q}{\sqrt{3}}\bigr)}\right)\nonumber\\
+ &\Bigl[\frac{2}{3}(1-z)\Bigr]^{-1/2-\nu_k}\left(\frac{C_1\,3^{-\Delta_-/2}\Gamma(\Delta_--1)\Gamma(2\nu_k)}
{\Gamma\bigl(\frac{\Delta_--1}{2}+\nu_k+\frac{q}{\sqrt{3}}\bigr)
\Gamma\bigl(\frac{\Delta_--1}{2}+\nu_k-\frac{q}{\sqrt{3}}\bigr)}
+\frac{C_2\,3^{-\Delta_+/2}\Gamma(\Delta_+-1)\Gamma(2\nu_k)}
{\Gamma\bigl(\frac{\Delta_+-1}{2}+\nu_k+\frac{q}{\sqrt{3}}\bigr)
\Gamma\bigl(\frac{\Delta_+-1}{2}+\nu_k-\frac{q}{\sqrt{3}}\bigr)}\right).
\end{align}
The asymptotic behavior near the boundary $z\to 0$ is
\begin{equation}
\phi\to C_1z^{\Delta_-}+C_2z^{\Delta_+}.
\end{equation}
By Eqs.~\eqref{eq:etah} and \eqref{eq:etab}, the solutions of $a_\pm^{(0)}$ and $b_\pm^{(0)}$ are
\begin{equation}
\begin{pmatrix}
a_+^{(0)} & a_-^{(0)}\\
b_+^{(0)} & b_-^{(0)}
\end{pmatrix}=
\frac{\nu_k}{\sqrt{m^2+4}}\begin{pmatrix}
\dfrac{18^{1/2+\nu_k}\cdot 3^{-\Delta_+/2}\Gamma(\Delta_+-1)\Gamma(2\nu_k)}
{\Gamma\bigl(\frac{\Delta_+-1}{2}+\nu_k+\frac{q}{\sqrt{3}}\bigr)
\Gamma\bigl(\frac{\Delta_+-1}{2}+\nu_k-\frac{q}{\sqrt{3}}\bigr)}
& \dfrac{-18^{1/2-\nu_k}\cdot 3^{-\Delta_+/2}\Gamma(\Delta_+-1)\Gamma(-2\nu_k)}
{\Gamma\bigl(\frac{\Delta_+-1}{2}-\nu_k+\frac{q}{\sqrt{3}}\bigr)
\Gamma\bigl(\frac{\Delta_+-1}{2}-\nu_k-\frac{q}{\sqrt{3}}\bigr)}\\
\dfrac{-18^{1/2+\nu_k}\cdot 3^{-\Delta_-/2}\Gamma(\Delta_--1)\Gamma(2\nu_k)}
{\Gamma\bigl(\frac{\Delta_--1}{2}+\nu_k+\frac{q}{\sqrt{3}}\bigr)
\Gamma\bigl(\frac{\Delta_--1}{2}+\nu_k-\frac{q}{\sqrt{3}}\bigr)}
& \dfrac{18^{1/2-\nu_k}\cdot 3^{-\Delta_-/2}\Gamma(\Delta_--1)\Gamma(-2\nu_k)}
{\Gamma\bigl(\frac{\Delta_--1}{2}-\nu_k+\frac{q}{\sqrt{3}}\bigr)
\Gamma\bigl(\frac{\Delta_--1}{2}-\nu_k-\frac{q}{\sqrt{3}}\bigr)}
\end{pmatrix}.\label{eq:a0b0}
\end{equation}
\end{widetext}
It can be checked that $a_+^{(0)}b_-^{(0)}-a_-^{(0)}b_+^{(0)}=\nu_k/\sqrt{m^2+4}$ is satisfied.

\section{Analytic Green's functions\label{sec:xi}}
\subsection{Zero modes and the phase diagram}
The zero mode is a gapless mode in the Green's function at $\omega=0$. The singularity of a fermionic Green's function at $\omega=0$ indicates a Fermi surface, while the singularity of a bosonic Green's function at $\omega=0$ indicates instability. In the fermionic case, solving $G^{-1}(\omega=0,\mathbf{k})=0$ gives $\mathbf{k}=\mathbf{k}_F$, where $\mathbf{k}_F$ is the Fermi momentum. Similarly, in the bosonic case, solving $G^{-1}(\omega=0,\mathbf{k})=0$ gives the $\mathbf{k}=\mathbf{k}_S$, where we use the subscript S for superfluid. The analytic solution enables us to solve for the zero modes, which are the only normal modes.

The zero modes are determined by $a_+^{(0)}=0$ for the standard quantization, and $b_+^{(0)}=0$ for the alternative quantization. By Eq.~\eqref{eq:a0b0}, we have
\begin{equation}
\frac{\Delta_\pm-1}{2}+\nu_k-\frac{q}{\sqrt{3}}=-n_\pm,\label{eq:nm}
\end{equation}
where $n_\pm$ is a nonnegative integer, the ``$+$" sign is for the standard quantization, and the ``$-$" sign is for the alternative quantization. Although the above result is derived under the assumption that $\Delta_+$ is not an integer, Eq.~\eqref{eq:nm} is generally valid when $m^2\geq -4$ for the standard quantization, and when $-4<m^2<-3$ for the alternative quantization (see Appendix~\ref{sec:notes}). The solution of $k$ is denoted by $k_S$. The existence of a real $k_S$ requires $\nu_k\geq 0$, which implies
\begin{equation}
\frac{q}{\sqrt{3}}\geq n_\pm+\frac{\Delta_\pm-1}{2}.
\end{equation}

The first type of instability is initiated by the nodeless zero mode, $n_\pm=0$. We set $n_\pm=0$ in the following. Solving $k$ from Eq.~\eqref{eq:nm} gives
\begin{align}
k_S^2 &=12\Bigl(\frac{q}{\sqrt{3}}-\frac{1\pm\sqrt{m^2+4}}{2}\Bigr)^2+2q^2-m^2-3\nonumber\\
&=6\Bigl[q-\frac{\sqrt{3}}{3}(1\pm\sqrt{m^2+4})\Bigr]^2+2(1\pm\sqrt{m^2+4}),\label{eq:kS}
\end{align}
where
\begin{equation}
q\geq\frac{\sqrt{3}}{2}(1\pm\sqrt{m^2+4}).\label{eq:qS}
\end{equation}
For the standard quantization, $k_S$ is always nonzero. For the alternative quantization, $k_S$ is nonzero except for a special case: $m^2=-3$ and $q=0$, in which, however, Eq.~\eqref{eq:sol} is not valid and the solution of $\phi$ by Eq.~\eqref{eq:marginal} below shows that $k=0$ is not a zero mode. Moreover, $k_S$ is always nonzero for all $n_\pm\geq 0$.

If Eq.~\eqref{eq:qS} is not satisfied, i.e., the zero mode does not exist, there can still be the second type of instability when $\nu_k$ is imaginary. The critical value of $k$ is
\begin{equation}
k_\text{IR}^2=2q^2-m^2-3,\label{eq:kIR}
\end{equation}
where
\begin{equation}
q^2>\frac{m^2+3}{2}.\label{eq:qIR}
\end{equation}
As explained in Ref.~\cite{Faulkner:2009wj}, this instability is due to the backreaction of pair productions in the IR geometry for a charged scalar, or the mass below the AdS$_2$ BF bound for a neutral scalar; the parameter set for which $\nu_k$ is imaginary is call the oscillatory region. The imaginary IR scaling dimension $\delta_k=1/2+\nu_k$ implies the conformality lost after the annihilation of two fixed points of the CFT dual to the AdS$_2$, leading to an instability of the IR geometry \cite{Kaplan:2009kr,Jensen:2010ga}.

\begin{figure*}
\begin{minipage}[t]{\textwidth}
%\centering
  \includegraphics[]{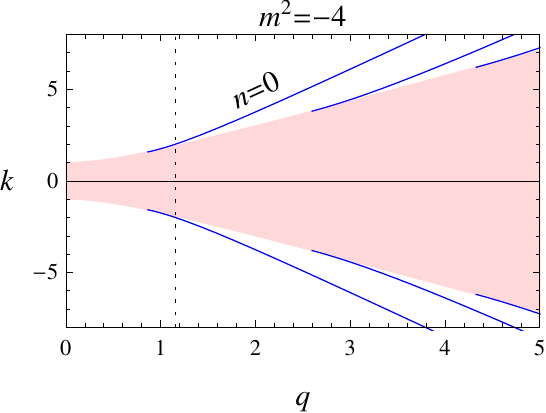}\quad
  \includegraphics[]{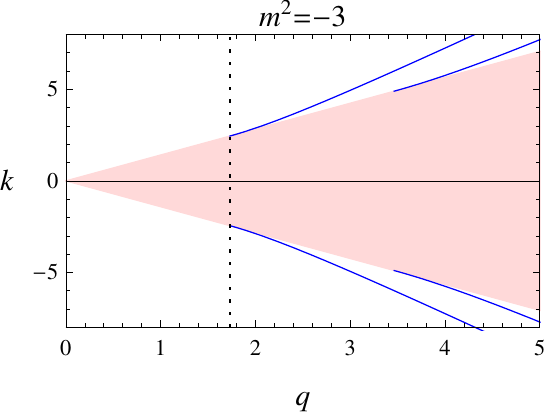}\quad
  \includegraphics[]{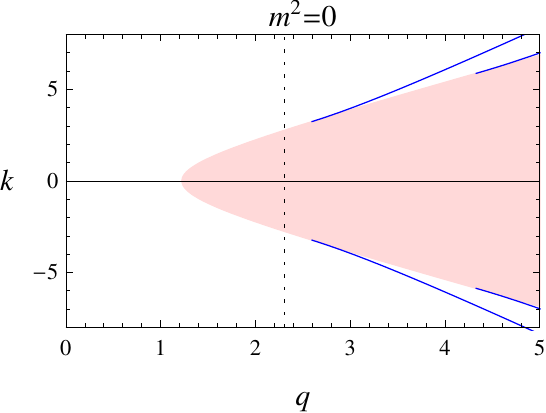}
  \caption{\label{fig:kqs} Phase diagram for the standard quantization. The solid lines correspond to zero modes. The oscillatory region is shaded, and will move to the right as we increase $m^2$. In the right plot, the tip of the oscillatory region corresponds to a bifurcating critical point at $k=0$. The dotted line is the BPS bound for $q\leq\Delta_+/\sqrt{3}$, where $q$ is the R-charge \cite{Denef:2009tp,Gubser:2009qm}.}
\end{minipage}\\[10pt]
\begin{minipage}[t]{\textwidth}
%\centering
  \includegraphics[]{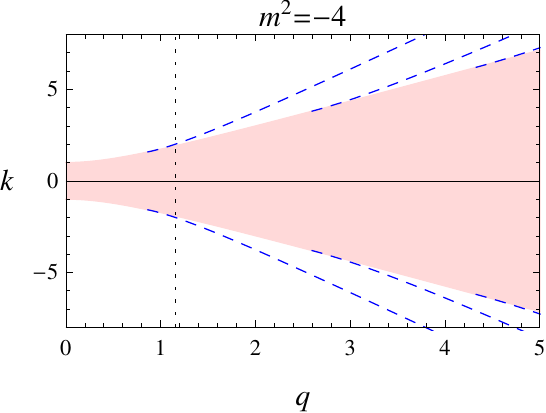}\quad
  \includegraphics[]{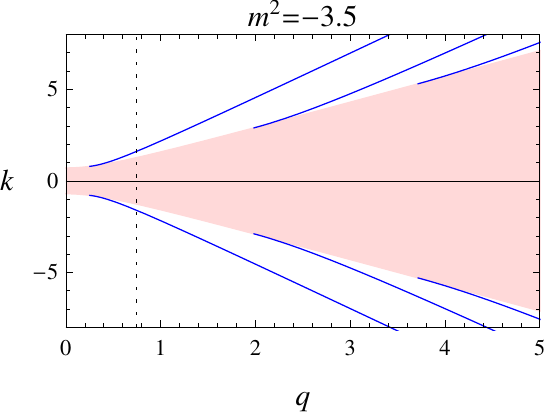}\quad
  \includegraphics[]{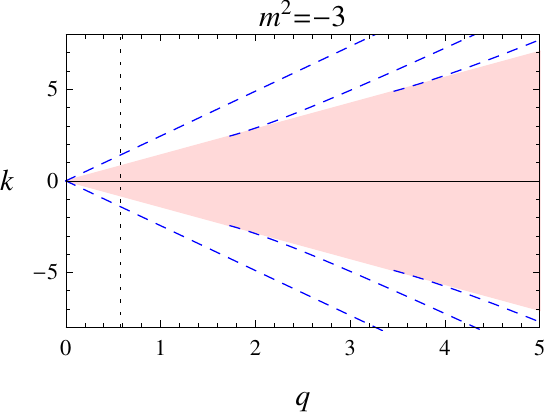}
  \caption{\label{fig:kqa} Phase diagram for the alternative quantization. The oscillatory region is shaded. The dashed lines are solutions to Eq.~\eqref{eq:nm}, but do not represent zero modes. The dotted line is the BPS bound for $q\leq\Delta_-/\sqrt{3}$.}
\end{minipage}
\end{figure*}

When a zero mode exists, there is always the second type of instability for $k<k_\text{IR}$, by comparing Eqs.~\eqref{eq:qS} and \eqref{eq:qIR}:
\begin{equation}
\sqrt{\frac{m^2+3}{2}}<\frac{\sqrt{3}}{2}(1\pm\sqrt{m^2+4}),
\end{equation}
and $k_\text{IR}\leq k_S$, by comparing Eqs.~\eqref{eq:kS} and \eqref{eq:kIR}. The solution of $k$ as a function of $m^2$ and $q$ is illustrated in Figs.~\ref{fig:kqs} and \ref{fig:kqa}. The blue lines are solutions of $k_S$ from Eq.~\eqref{eq:nm}; the outermost one is $n=0$. The boundary of the shaded region is the solution of $k_\text{IR}$ from Eq.~\eqref{eq:kIR}.

\begin{table*}
  \caption{\label{tab:sum} The critical value of the superfluid velocity ($\pm$ is for the standard and alternative quantization, respectively)}
  \begin{ruledtabular}
  \begin{tabular}{lll}
  $m^2$ & $q$ & Superfluid velocity $S_x$\\\hline
  \multirow{2}{*}{$-4\leq m^2\leq -3$\,\footnote{For the alternative quantization, the $m^2=-4$ and $-3$ cases are not included.}} & $0\leq q\leq\frac{\sqrt{3}}{2}(1\pm\sqrt{m^2+4})$ & IR geometry instability at $S_x=k_\text{IR}/q$\\
  & $q\geq\frac{\sqrt{3}}{2}(1\pm\sqrt{m^2+4})$ & Zero mode instability at $S_x=k_S/q$\\\hline
  \multirow{3}{*}{$m^2>-3$} & $0\leq q<\sqrt{\frac{m^2+3}{2}}$ & No instability\\
  & $\sqrt{\frac{m^2+3}{2}}\leq q\leq\frac{\sqrt{3}}{2}(1+\sqrt{m^2+4})$ & IR geometry instability at $S_x=k_\text{IR}/q$\\
  & $q\geq\frac{\sqrt{3}}{2}(1+\sqrt{m^2+4})$ & Zero mode instability at $S_x=k_S/q$
  \end{tabular}
  \end{ruledtabular}
\end{table*}

By perturbation around the solution at $\omega=0$, we can obtain the retarded Green's function near the zero mode
\begin{equation}
G=\frac{h_1}{k_\perp-\frac{1}{v_S}\omega-h_2e^{i\gamma_{k_S}}\omega^{2\nu_{k_S}}},
\end{equation}
where $k_\perp=k-k_S$, and the quantities $v_S$, $h_1$, and $h_2$ can be calculated by the formulas given in Appendix C of Ref.~\cite{Faulkner:2009wj}. As explained in Ref.~\cite{Faulkner:2009wj}, when the momentum changes from $k>k_S$ to $k<k_S$, a pole of the Green's function moves across the origin to the upper half complex $\omega$-plane, signaling an instability. As a comparison, analytic fermionic Green's functions can be obtained for a massless spinor in the background of the two-charge black hole in AdS$_5$; the Fermi momenta are $k_F=q-n-1/2$ (in units of the chemical potential), where $q$ is the charge of the spinor, and $n$ labels the Fermi surface \cite{Gubser:2012yb}. The RN-AdS black hole background is more complicated due to the extra feature of the oscillatory region.

\begin{figure}
  \centering
  \includegraphics[]{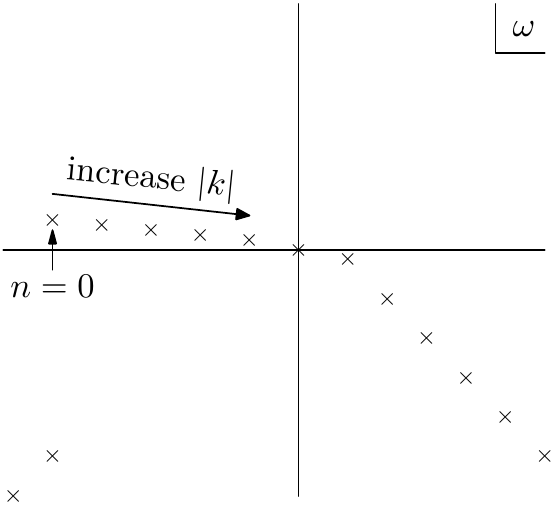}
  \caption{\label{fig:qnm} Schematic plot of the poles of the Green's function when $q$ is large.}
\end{figure}

It is helpful to understand the normal modes by looking at the poles of the Green's function at arbitrary $\omega$. The numerical calculations suggest the following features, as illustrated in Fig.~\ref{fig:qnm}. At $q=0$ and $k=0$, all the poles of the Green's function are in the lower half $\omega$-plane, and not close to $\omega=0$. As we increase the charge $q$, there are more and more poles moving across the origin to the upper half $\omega$-plane, and the first one is labeled by $n=0$. Suppose we start from a large $q$. As we increase $k$, the poles will move to the right, across the origin to the lower half $\omega$-plane. Therefore, the largest $k$ corresponds to the $n=0$ mode, which triggers the onset of the instability. The situation is similar to the fermionic case with a crucial difference that the quasibound states in the upper half plane in Fig.~\ref{fig:qnm} are now in the lower half plane \cite{Herzog:2012kx}.

According to Ref.~\cite{Iqbal:2011aj}, we can associate the first type of instability with the hybridized critical point, and the second type of instability with the bifurcating critical point. Recall that the system has rotational symmetry. In the fermionic case, a nonzero $k_F$ gives a spherical Fermi surface. However, in the bosonic case, the condensation of the scalar cannot happen at a spherical shell of the momentum space. Instead, the quantum phase transition happens at $\omega=0$ and $k=0$. Without tuning extra parameters, only the second type of instability can happen. By tuning an extra parameter, both types of instability can happen at $k=0$.

\subsection{Bifurcating critical point\label{sec:xib}}
When the IR scaling exponent $\nu_k$ at $k=0$ becomes imaginary, there is a bifurcating critical point \cite{Iqbal:2011aj}. We will denote $\nu:=\nu_{k=0}=\frac{1}{2\sqrt{3}}\sqrt{u}$, where $u=m^2-2q^2+3$. From $u>0$ to $u<0$, the Green's function at $\omega=0$ keeps finite, but bifurcates into the complex plane and has a cusp \cite{Iqbal:2011aj}.

Near $\nu_k=0$, the quantities $a_\pm^{(0)}$ and $b_\pm^{(0)}$ in the Green's function can be expanded as
\begin{align}
a_\pm^{(0)} &=\alpha\pm\nu_k\tilde{\alpha}+\cdots,\nonumber\\
b_\pm^{(0)} &=\beta\pm\nu_k\tilde{\beta}+\cdots.
\end{align}
By Eq.~\eqref{eq:a0b0}, we have
\begin{align}
\alpha &=\tfrac{3^{1-\Delta_+/2}\Gamma(\Delta_+-1)}
{\sqrt{2(m^2+4)}\,\Gamma\bigl(\frac{\Delta_+-1}{2}+\frac{q}{\sqrt{3}}\bigr)
\Gamma\bigl(\frac{\Delta_+-1}{2}-\frac{q}{\sqrt{3}}\bigr)}\nonumber\\
\tilde{\alpha} &=-\tfrac{\Gamma(\Delta_+-1)\bigl(\psi\bigl(\frac{\Delta_+-1}{2}+\frac{q}{\sqrt{3}}\bigr)
+\psi\bigl(\frac{\Delta_+-1}{2}-\frac{q}{\sqrt{3}}\bigr)+2\gamma-\ln 18\bigr)}
{\sqrt{2(m^2+4)}\,3^{-1+\Delta_+/2}\Gamma\bigl(\frac{\Delta_+-1}{2}+\frac{q}{\sqrt{3}}\bigr)
\Gamma\bigl(\frac{\Delta_+-1}{2}-\frac{q}{\sqrt{3}}\bigr)}\nonumber\\
\beta &=-\tfrac{3^{1-\Delta_-/2}\Gamma(\Delta_--1)}
{\sqrt{2(m^2+4)}\,\Gamma\bigl(\frac{\Delta_--1}{2}+\frac{q}{\sqrt{3}}\bigr)
\Gamma\bigl(\frac{\Delta_--1}{2}-\frac{q}{\sqrt{3}}\bigr)}\nonumber\\
\tilde{\beta} &=\tfrac{\Gamma(\Delta_--1)\bigl(\psi\bigl(\frac{\Delta_--1}{2}+\frac{q}{\sqrt{3}}\bigr)
+\psi\bigl(\frac{\Delta_--1}{2}-\frac{q}{\sqrt{3}}\bigr)+2\gamma-\ln 18\bigr)}
{\sqrt{2(m^2+4)}\,3^{-1+\Delta_-/2}\Gamma\bigl(\frac{\Delta_--1}{2}+\frac{q}{\sqrt{3}}\bigr)
\Gamma\bigl(\frac{\Delta_--1}{2}-\frac{q}{\sqrt{3}}\bigr)},\label{eq:alphabeta}
\end{align}
where $\psi(x):=\Gamma'(x)/\Gamma(x)$ is the digamma function, and $\gamma$ is the Euler-Mascheroni constant. It can be checked that $\alpha\tilde{\beta}-\beta\tilde{\alpha}=-1/(2\sqrt{m^2+4})$ is satisfied.

Near the bifurcating critical point $\nu\to 0$, the IR Green's function can be written as
\begin{equation}
\mathcal{G}_{k=0}(\omega)=-1+2\nu\mathcal{G}_0(\omega),
\end{equation}
where $\mathcal{G}_0(\omega)$ is the IR Green's function when $\nu=0$, which can be obtained by Eq.~\eqref{eq:phiin} with $\nu_k=0$:
\begin{equation}
\mathcal{G}_0(\omega)=-\ln(-2i\omega)-2\gamma-\psi\Bigl(\frac{1}{2}-\frac{iq}{\sqrt{6}}\Bigr).
\end{equation}
Particularly, for a neutral scalar,
\begin{equation}
\mathcal{G}_0(\omega)=-\ln\Bigl(-\frac{i\omega}{2}\Bigr)-\gamma.\label{eq:gir00}
\end{equation}
By expanding the Green's function, Eq.~\eqref{eq:green}, at small $\nu$, we obtain the Green's function for the bifurcating critical point:
\begin{equation}
G=\frac{\beta\,\mathcal{G}_0(\omega)+\tilde{\beta}}{\alpha\,\mathcal{G}_0(\omega)+\tilde{\alpha}}.\label{eq:gbifur}
\end{equation}
The finite temperature generalization of $\mathcal{G}_0(\omega)$ is given by Eqs.~\eqref{eq:gir0T} and \eqref{eq:gir00T} in Appendix~\ref{sec:gir}.

Note that Eq.~\eqref{eq:gbifur} is obtained by taking the $\nu\to 0$ limit at fixed $\omega$. If we want to examine the poles of the Green's function, we need a Green's function valid to arbitrarily small $\omega$. To do so, we need to use $\mathcal{G}_{k=0}(\omega)$ instead of $\mathcal{G}_0(\omega)$. In the condensed side ($u<0)$, there are infinite number of poles in the upper half $\omega$-plane; these massive states will condense and lead to instability \cite{Iqbal:2011aj}.

\subsection{Critical points with superfluid velocity\label{sec:xic}}
A superfluid with a supercurrent flow can be studied holographically by turning on a vector potential
\begin{equation}
A_x=S_x+J^xz^2+\cdots,
\end{equation}
where $S_x$ is the superfluid velocity, which is the source, and $J^x$ is the supercurrent, which is the expectation value \cite{Basu:2008st,Herzog:2008he,Sonner:2010yx,Arean:2010zw,Arean:2010wu}. The Klein-Gordon equation for a scalar $\phi$ coupled with both $A_t$ and $A_x$ at $\omega=0$ and $k=0$ is
\begin{equation}
\phi''+\left(\frac{f'}{f}-\frac{3}{z}\right)\phi'
+\left(\frac{q^2A_t^2}{f^2}-\frac{q^2A_x^2}{f}-\frac{m^2}{z^2f}\right)\phi=0.\label{eq:phiA}
\end{equation}
When the superfluid velocity is too large, the superfluid phase will return to the normal phase. At the critical point, $A_x=S_x$ is a constant. From Eq.~\eqref{eq:phiA}, we can see that the superfluid velocity plays the same role as the momentum $k$,\footnote{We reinterpret $k$ as the superfluid velocity (in the $x$-direction). Another way to reinterpret $k$ is the magnetic field added by $A_y=Bx$, assuming that the metric is not changed. After a separation of variables $\Phi\sim X(x)\phi(z)$, the magnetic filed plays the same role as $k^2$ ($qB\leftrightarrow k^2$) in the equation for $\phi$ \cite{Albash:2008eh,Hartnoll:2008kx}. In terms of a dyonic black hole in AdS$_4$, the quantum critical point by the second type of instability is studied in Ref.~\cite{Iqbal:2010eh}.} which is consistent with the fact that the superfluid velocity is the gradient of the phase of the order parameter. The phase diagrams are Figs.~\ref{fig:kqs} and \ref{fig:kqa} by replacing $k$ with $qS_x$.

According to previous studies of holographic superconductors, there are two mechanisms for instabilities: (1) When the charge of the scalar is large, the effective mass $m_\text{eff}^2=m^2+g^{tt}q^2A_t^2$ is negative enough to produce an unstable mode \cite{Gubser:2008px,Hartnoll:2008vx}. (2) When it is close to the zero temperature, and the effective mass $m_\text{eff}^2=(m^2-2q^2)/12$ is below the AdS$_2$ BF bound $m_\text{BF}^2=-1/4$, the IR geometry is unstable \cite{Hartnoll:2008kx,Faulkner:2009wj,Horowitz:2009ij}. Here the quantum critical point can be reached by tuning the superfluid velocity, and the above two mechanisms for instabilities correspond to the two types of instabilities we discussed before: zero mode instability and IR geometry instability, respectively. They can happen at the same time, giving a mixed critical point, as shown in Figs.~\ref{fig:kqs} and \ref{fig:kqa}, when a solid line touches the oscillatory region.

We call the stable region ($k$ or $qS_x>\max(k_S,k_\text{IR})$) the normal phase, and we call the unstable region the superfluid phase. When the mass is small ($-4\leq m^2\leq -3$), there are two cases: (i) For small charge, the scalar is unstable due to the second type of instability; (ii) For large charge, the scalar is unstable due to the first type of instability. When the mass is large ($m^2>-3$), there are three cases: (i) For small charge, the scalar is stable; (ii) For intermediate charge, the scalar is unstable due to second type of instability; (iii) For large charge, the scalar is unstable due to the first type of instability. The result is summarized in Table~\ref{tab:sum}.

\section{Adding a double trace deformation\label{sec:kappa}}
\subsection{Hybridized critical point}
\begin{figure}
  \centering
  \includegraphics[]{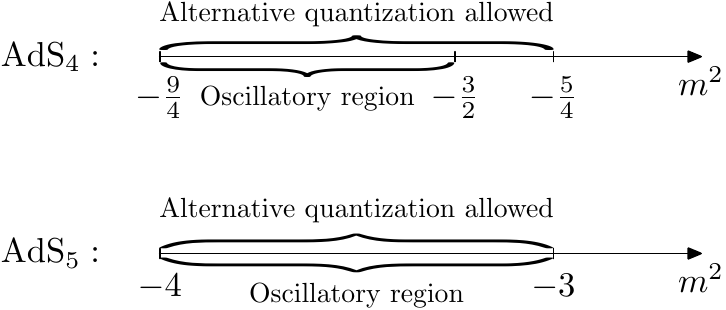}
  \caption{\label{fig:m2osc} A slight difference between AdS$_4$ and AdS$_5$. We set the AdS radius $L=1$. In AdS$_5$, we can only start from the standard quantization.}
\end{figure}

Without the superfluid velocity, the zero mode at $k=0$ can also be achieved by tuning another parameter $\kappa_+$, which describes a double trace deformation in the boundary CFT:
\begin{equation}
\frac{\kappa_+}{2}\int d^dx\,\mathcal{O}^2,\label{eq:deform}
\end{equation}
where $\langle\mathcal{O}\rangle=B$ \cite{Iqbal:2011aj,Witten:2001ua}. The Green's function becomes
\begin{equation}
G^{(\kappa_+)}=\frac{1}{G^{-1}+\kappa_+}.\label{eq:Gkappa}
\end{equation}
Similarly, starting from the alternative quantization, we can add a double trace deformation with coefficient $\kappa_-$, which is related to $\kappa_+$ by $\kappa_-=-1/\kappa_+$. The alternative quantization is allowed only if the $m^2$ of the scalar is in the interval $m_\text{BF}^2\leq m^2\leq m_\text{BF}^2+1$ \cite{Klebanov:1999tb}, which leads to a slight difference between AdS$_4$ and AdS$_5$ as illustrated in Fig.~\ref{fig:m2osc}. Since we are interested in the instability triggered by a zero mode at $k=0$, we do not want other instabilities to exist. Therefore, we require $u>0$, i.e., the parameters are not in the oscillatory region, in which the system is already unstable. In AdS$_4$, if we start from the alternative quantization, there is still an interval of $m^2$ not in the oscillatory region. In AdS$_5$, however, we can only start from the standard quantization.\footnote{When $m^2\geq -3$, Eq.~\eqref{eq:deform} is an irrelevant term. We assume that the UV geometry is not changed, and examine (both UV and IR) instabilities by the Green's function.}

The Green's function at the leading order in $\omega$ is
\begin{equation}
G(\omega,k)=\frac{b_+^{(0)}+{\cal G}_k(\omega)b_-^{(0)}}
{a_+^{(0)}+\kappa_+b_+^{(0)}+{\cal G}_k(\omega)(a_-^{(0)}+\kappa_+b_-^{(0)})}.\label{eq:Ghybri}
\end{equation}
The boundary condition for a pole in the Green's function at $\omega=0$ is $a_+^{(0)}+\kappa_+b_+^{(0)}=0$. Thus, by Eq.~\eqref{eq:a0b0}, the critical value of $\kappa_+$ is
\begin{multline}
\kappa_c=3^{-\sqrt{m^2+4}}\\
\times\frac{\Gamma\bigl(\frac{\Delta_--1}{2}+\nu_k+\frac{q}{\sqrt{3}}\bigr)
\Gamma\bigl(\frac{\Delta_--1}{2}+\nu_k-\frac{q}{\sqrt{3}}\bigr)\Gamma(\Delta_+-1)}
{\Gamma\bigl(\frac{\Delta_+-1}{2}+\nu_k+\frac{q}{\sqrt{3}}\bigr)
\Gamma\bigl(\frac{\Delta_+-1}{2}+\nu_k-\frac{q}{\sqrt{3}}\bigr)\Gamma(\Delta_--1)}.\label{eq:kappa}
\end{multline}
When $\kappa_+=\kappa_c$, we obtain a hybridized critical point, which is described by an order parameter in the Ginzburg-Landau sector hybridized with a strongly coupled sector, the emergent CFT$_1$ dual to the IR AdS$_2$ \cite{Iqbal:2011aj}. The Green's function near a hybridized critical point can be written as
\begin{equation}
G(\omega,k)=\frac{1}{\kappa_+-\kappa_c+h_kk^2-h_\omega\omega+hC(\nu)(-i\omega)^{2\nu}},\label{eq:ghybri}
\end{equation}
where
\begin{align}
h &=\frac{a_-^{(0)}+\kappa_cb_-^{(0)}}{b_+^{(0)}}
=-\frac{\nu}{\sqrt{m^2+4}\,(b_+^{(0)})^2}\nonumber\\
C &=\frac{\Gamma(-2\nu)\Gamma\bigl(\frac{1}{2}+\nu-\frac{iq}{\sqrt{6}}\bigr)}
{\Gamma(2\nu)\Gamma\bigl(\frac{1}{2}-\nu-\frac{iq}{\sqrt{6}}\bigr)}2^{2\nu}.
\end{align}
For a neutral scalar, the first order in $\omega$ vanishes, so we need to write $h_\omega\omega^2$ instead.

We assume $\nu<1/2$ for a charged scalar, or $\nu<1$ for a neutral scalar. For example, $\nu<1$ for a neutral scalar means $-3<m^2<9$, which is already a large range. Then we do not need $h_\omega$ since the $\omega^{2\nu}$ term is dominant as $\omega\to 0$. For the Green's function, $\omega=0$ is a branch point in the complex $\omega$-plane, and we define the physical sheet as $\theta\in(-\pi/2,3\pi/2)$. The Green's function has a pole at
\begin{equation}
\omega_*=i\left(\frac{\kappa_c-\kappa_+}{hC(\nu)}\right)^\frac{1}{2\nu}.
\end{equation}
Note that $h<0$, and for $q>0$,
\begin{equation}
0<\arg\frac{1}{(-C)^{1/2\nu}}<\frac{\pi}{2},\qquad \nu<\frac{1}{2},\label{eq:C}
\end{equation}
(see Appendix~\ref{sec:gir}); for $q=0$, $C$ is real and negative. Therefore, when $\kappa_+<\kappa_c$, the pole will be in the upper half $\omega$-plane of the physical sheet. When $\kappa_+>\kappa_c$, the pole will be either in the lower half $\omega$-plane of the physical sheet, or on a non-physical sheet.

\begin{figure*}
  %\centering
  \includegraphics[]{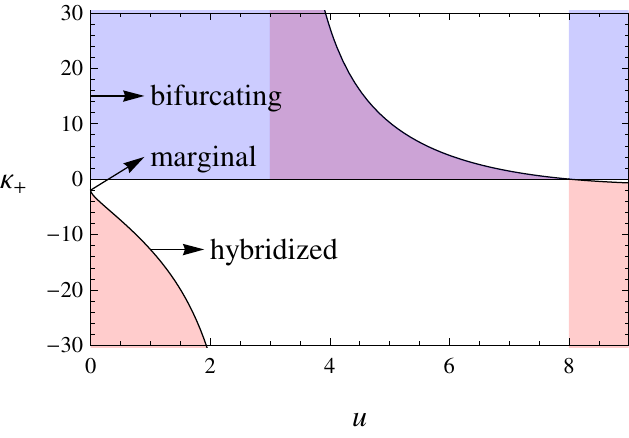}\qquad
  \includegraphics[]{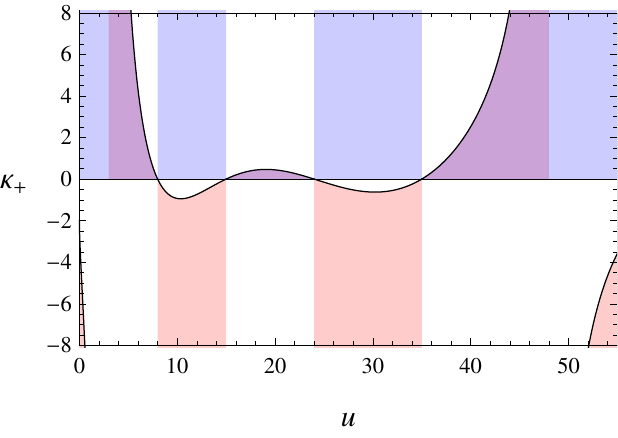}
  \caption{\label{fig:pdiag12} The tentative phase diagram for a neutral scalar ($q=0$), with different ranges of $u=m^2+3$. The small $u$ part is consistent with Ref.~\cite{Iqbal:2011aj}. The bifurcating critical point is at $u=0$, and the oscillatory region is $u<0$. The hybridized critical point is at $\kappa_+=\kappa_c$. The curve crosses the horizontal axis at $u=8$. The red shaded region when $\kappa_+<\kappa_c$ is for IR instability, and the blue shaded region when $\kappa_+>0$ is for UV instability. The white regions are stable.}
\end{figure*}

We can easily plot the critical value of $\kappa_+$ for the hybridized critical point as a function of $u$, as shown in Fig.~\ref{fig:pdiag12}. The boundary for the bifurcating critical point is $u=0$. The two curves $u=0$ and $\kappa_+=\kappa_c$ intersect at $(u,\kappa_+)=(0,-2)$, which is a marginal critical point. When $\kappa_+$ changes across the curve from $\kappa_+>\kappa_c$ to $\kappa_+<\kappa_c$, a pole will move across the origin to the upper half $\omega$-plane, causing an instability. We can see that $\kappa_c$ is a single-valued function of $u$, which implies the following. If we keep increasing or decreasing $\kappa_+$, after a pole moves across the origin of the complex $\omega$-plane, it will never come back to the origin. However, we cannot exclude the possibility that a pole can appear from infinity in the upper half $\omega$-plane, or a pole in the upper half $\omega$-plane can move to infinity and disappear. The analytic solution cannot capture these features. We hope further numerical calculations can be helpful to make this clear.\footnote{There are two challenges in numerical calculations. One is that the result becomes inaccurate very quickly as the $\textrm{Im}(\omega)$ becomes large; the other is that it is more difficult to obtain the expectation value accurately when the scaling dimension is large.}

We will refer to the IR instability as the singularity of the Green's function at $|\omega|\to 0$, and UV instability as the singularity of the Green's function at $|\omega|\to\infty$ in the upper half complex $\omega$-plane. For a CFT at zero density, the boundary for the UV instability is $\kappa_+=0$ \cite{Iqbal:2011aj}. If the UV and IR instabilities are related, a tentative phase diagram for a neutral scalar is shown in Fig.~\ref{fig:pdiag12}. The boundary for the hybridized critical point is much more intricate than previous numerical results showed \cite{Iqbal:2011aj}. Especially, the curve will cross the $u$ axis ($\kappa_+=0$). From Eq.~\eqref{eq:kappa}, we can see that when $1-\Delta_-=-n$, i.e., $m^2=(n+1)^2-4$, where $n=0$, $1$, $\cdots$, the denominator of $\kappa_c$ is infinity; thus, if there is no other infinity in the numerator, we have $\kappa_c=0$, which implies that the static Green's function $G(\omega=0)\to\infty$. However, when $\Delta_+$ (or $\Delta_-$) is a integer, we need to use Eq.~\eqref{eq:sol2} in Appendix~\ref{sec:notes} to calculate the Green's function, and the result is usually finite. This puzzle is largely related to the renormalization near integer values of $\Delta_\pm$ (see Appendix~\ref{sec:notes}).

\subsection{Marginal critical point}
The marginal critical point happens at both $u=0$ and $\kappa_+=\kappa_c$. Recall that the IR scaling dimension $\delta_k=1/2+\nu_k$; when $\nu=0$, the double trace deformation is marginally relevant in the IR CFT \cite{Iqbal:2011aj}. The only possible parameters for the marginal critical point are $m^2=-3$ and $q=0$. The critical value of $\kappa_+$ is $\kappa_c=-2$, which can be obtained from Eq.~\eqref{eq:kappa} by setting $q=0$ and $k=0$ first, and then taking the $m^2\to -3$ limit. The direct calculations are as follows. The solution of $\phi$ in the outer region at $\omega=0$ and $k=0$ is
\begin{equation}
\phi_O=\frac{z}{\sqrt{1-z^2}}\left(C_1+C_2\ln\frac{1-z^2}{2z^2+1}\right).\label{eq:marginal}
\end{equation}
The solution in the inner region is
\begin{equation}
\phi_I=\frac{\ln[12(1-z)]}{\sqrt{12(1-z)}}+\mathcal{G}_0(\omega)\frac{1}{\sqrt{12(1-z)}},
\end{equation}
where $\mathcal{G}_0$ is given by Eq.~\eqref{eq:gir00}. By matching $\phi_O$ and $\phi_I$, we obtain
\begin{align}
\alpha &=\frac{1}{\sqrt{6}} &\quad \beta &=\frac{1}{2\sqrt{6}}\nonumber\\
\tilde{\alpha} &=\frac{1}{\sqrt{6}}\ln 18 &\quad \tilde{\beta} &=\frac{1}{2\sqrt{6}}\ln 18-\frac{3}{\sqrt{6}}.
\end{align}
By expanding Eq.~\eqref{eq:Ghybri} at small $\nu$, the Green's function near the marginal critical point is
\begin{equation}
G=\frac{\beta\,\mathcal{G}_0(\omega)+\tilde{\beta}}
{(\alpha+\kappa_+\beta)\mathcal{G}_0(\omega)+\tilde{\alpha}+\kappa_+\tilde{\beta}},
\end{equation}
which gives
\begin{equation}
G=\frac{\ln\omega-2\ln 6+6+\gamma-i\pi/2}{(\kappa_++2)(\ln\omega-2\ln 6+6+\gamma-i\pi/2)-12},\label{eq:gmar}
\end{equation}
where the critical value of $\kappa_+$ is $-2$.

We can replace the zero temperature Green's function $\mathcal{G}_k(\omega)$ with the finite temperature Green's function, Eq.~\eqref{eq:girT} in Appendix~\ref{sec:gir}, and the phase diagram of $T$-$\kappa_+$ can be obtained \cite{Iqbal:2011aj}. At finite temperature $T<<\mu$, the result is
\begin{equation}
G=\frac{\psi\bigl(\frac{1}{2}-\frac{i\omega}{2\pi T}\bigr)+\ln\frac{\pi T}{18}+6+\gamma}
{(\kappa_++2)\bigl[\psi\bigl(\frac{1}{2}-\frac{i\omega}{2\pi T}\bigr)+\ln\frac{\pi T}{18}+6+\gamma\bigr]-12},
\end{equation}
whose imaginary part is
\begin{equation}
\text{Im}G=\frac{\pi}{24}\tanh\frac{\omega}{2T}=\begin{cases}
\dfrac{\pi\omega}{48T}\quad &\omega<<T\\[8pt]
\dfrac{\pi}{24}\text{sgn}(\omega)\quad &\omega>>T.
\end{cases}
\end{equation}

\section{Discussion\label{sec:sum}}
Starting from the extremal RN-AdS$_5$ black hole, we have studied the quantum critical points by solving the Klein-Gordon equation in the bulk. The result gives us a glimpse of some strongly interacting systems, whose properties are beyond the reach of the perturbative method in quantum field theory. Just like the harmonic oscillator and the hydrogen atom as exactly solvable models capture essential features in quantum mechanics, the exactly solvable model in this work, together with a previous fermionic one \cite{Gubser:2012yb}, captures many essential features in AdS/CMT.

\begin{figure}
  \centering
  \includegraphics[]{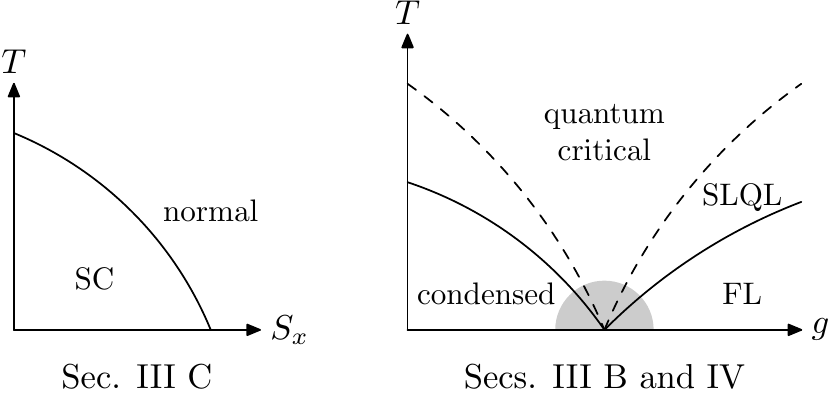}
  \caption{\label{fig:pdiagT} The left plot is the phase diagram for Sec.~\ref{sec:xic}. The parameter $S_x$ is the superfluid velocity. SC denotes (holographic) superconductor. The right plot is the phase diagram for Secs.~\ref{sec:xib} and \ref{sec:kappa}, according to Ref.~\cite{Iqbal:2011aj}. The parameter $g$ is $u$ for the bifurcating critical point, or $\kappa_+$ for the hybridized critical point.}
\end{figure}

We have calculated the Green's function to the leading order in $\omega$. The non-analyticity of the Green's function indicates two types of instabilities: one is triggered by a zero mode, and the other is triggered by the instability of the IR geometry. In the standard/alternative quantization, the zero modes of the system are always at finite $\mathbf{k}$. However, we can tune an extra parameter to make the zero mode be at $\mathbf{k}=0$. We considered the quantum critical points of two systems, whose finite temperature phase diagrams are illustrated in Fig.~\ref{fig:pdiagT}. The extra parameter in the first system is the superfluid velocity. The RN-AdS$_5$ geometry describes the normal phase, and depending on the parameters $(m^2,q)$, the system can develop zero mode instability, IR geometry instability, or be stable. The extra parameter in the second system is the double trace deformation. The zero mode instability gives a hybridized critical point. In the second system, besides the above IR instabilities, there is also UV instability, which is not captured by the analytic solution.

There are several remaining questions in the second system as follows. (i) What is the full phase diagram including both UV and IR instabilities? We already have the boundary for the IR instabilities. To find the stable region, we need not only the boundary for the UV instabilities, but how the poles move as we change the parameters. (ii) What happens to the Green's function when $\Delta_+$ approaches an integer? This is related to the curve crossing the $u$-axis ($\kappa_+=0$) in Fig.~\ref{fig:pdiag12}. It seems that the $\Delta_+$ approaching an integer limit is not the same as the result obtained by setting $\Delta_+$ be the integer. (iii) What are the analogous models in condensed matter physics?

Note added: Recently there is another related work \cite{Alishahiha:2012ad} for the two-charge black hole in AdS$_5$, in which the massless Klein-Gordon equation is analytically solved. We can obtain that the zero modes as $\nu_k=q-2n-3$, where $\nu_k=\sqrt{k^2+1}$ and $n$ is a nonnegative integer. Only the first type of instability exists, because the electric field approaches to zero in the near horizon limit \cite{Gubser:2012yb}.

\acknowledgments
I thank Profs. S.S. Gubser, F.D.M. Haldane, C.P. Herzog, and D.A. Huse for helpful discussions, and thank N. Iqbal for communications. I also thank YITP and SCGP
%\textit{C.N. Yang Institute for Theoretical Physics} and \textit{Simons Center for Geometry and Physics}
at Stony Brook for their hospitalities during the completion of this work. This work was supported in part by the Department of Energy under Grant No.~DE-FG02-91ER40671, and the National Science Foundation under Grants No.~PHY-0844827 and PHY-0756966.

\appendix
\section{Mathematical notes\label{sec:notes}}
\textit{Hypergeometric function ${_2F_1}(\alpha,\beta;\gamma;x)$.} We will denote ${_2F_1}$ by $F$ for simplicity. The following formulas are helpful to obtain and understand the analytic solutions:
\begin{equation}
F(\alpha,\beta;\gamma;x)=(1-x)^{-\beta}F\bigl(\gamma-\alpha,\beta;\gamma;\frac{x}{x-1}\bigr).\label{eq:trans}
\end{equation}
If $\text{Re}(\gamma)>\text{Re}(\alpha+\beta)$,
\begin{equation}
F(\alpha,\beta;\gamma;1)=\frac{\Gamma(\gamma)\Gamma(\gamma-\alpha-\beta)}{\Gamma(\gamma-\alpha)\Gamma(\gamma-\beta)}.
\end{equation}
A connection formula for the hypergeometric function is
\begin{multline}
F(\alpha,\beta;\gamma;x)=AF(\alpha,\beta;\alpha+\beta-\gamma+1;1-x)\\
+B(1-x)^{\gamma-\alpha-\beta}F(\gamma-\alpha,\gamma-\beta;\gamma-\alpha-\beta+1;1-x),
\end{multline}
where
\begin{equation}
A=\frac{\Gamma(\gamma)\Gamma(\gamma-\alpha-\beta)}{\Gamma(\gamma-\alpha)\Gamma(\gamma-\beta)},\qquad
B=\frac{\Gamma(\gamma)\Gamma(\alpha+\beta-\gamma)}{\Gamma(\alpha)\Gamma(\beta)}.
\end{equation}

\textit{Other solutions for the Klein-Gordon equation~\eqref{eq:KG}.} If $\Delta_+$ is an integer, Eq.~\eqref{eq:sol} is no longer a general solution. If $2\nu_k$ is not an integer, the general solution for Eq.~\eqref{eq:KG} at $\omega=0$ can be written as
\begin{widetext}
\begin{align}
\phi(z)= &C_1\frac{z^{\Delta_+}(1-z^2)^{-1/2-\nu_k}}{(2z^2+1)^{-1/2-\nu_k+\Delta_+/2}}\,
{_2F_1}\Bigl(\frac{\Delta_+-1}{2}-\nu_k-\frac{q}{\sqrt{3}},\,\frac{\Delta_+-1}{2}-\nu_k+\frac{q}{\sqrt{3}};\,
1-2\nu_k;\,\frac{1-z^2}{2z^2+1}\Bigr)\nonumber\\
+ &C_2\frac{z^{\Delta_+}(1-z^2)^{-1/2+\nu_k}}{(2z^2+1)^{-1/2+\nu_k+\Delta_+/2}}\,
{_2F_1}\Bigl(\frac{\Delta_+-1}{2}+\nu_k-\frac{q}{\sqrt{3}},\,\frac{\Delta_+-1}{2}+\nu_k+\frac{q}{\sqrt{3}};\,
1+2\nu_k;\,\frac{1-z^2}{2z^2+1}\Bigr).\label{eq:sol2}
\end{align}
If $2q/\sqrt{3}$ is not an integer, the general solution can also be written as
\begin{align}
\phi(z)= &C_1z^{\Delta_+}\frac{(1-z^2)^{-\Delta_+/2+q/\sqrt{3}}}{(2z^2+1)^{q/\sqrt{3}}}\,
{_2F_1}\Bigl(\frac{\Delta_+-1}{2}-\nu_k-\frac{q}{\sqrt{3}},\,\frac{\Delta_+-1}{2}+\nu_k-\frac{q}{\sqrt{3}};\,
1-\frac{2q}{\sqrt{3}};\,\frac{2z^2+1}{1-z^2}\Bigr)\nonumber\\
+ &C_2z^{\Delta_+}\frac{(1-z^2)^{-\Delta_+/2-q/\sqrt{3}}}{(2z^2+1)^{-q/\sqrt{3}}}\,
{_2F_1}\Bigl(\frac{\Delta_+-1}{2}-\nu_k+\frac{q}{\sqrt{3}},\,\frac{\Delta_+-1}{2}+\nu_k+\frac{q}{\sqrt{3}};\,
1+\frac{2q}{\sqrt{3}};\,\frac{2z^2+1}{1-z^2}\Bigr).\label{eq:sol3}
\end{align}
\end{widetext}
If all $\Delta_+$, $2\nu_k$, and $2q/\sqrt{3}$ are integers, it is likely that the solution is an elementary function. It is especially convenient to use Eq.~\eqref{eq:sol2}, because the boundary condition at the horizon to obtain zero modes is simply $C_1=0$. We will discuss some special cases, in which we only consider the standard quantization.

\textit{Special case 1}: $m^2=-4$ ($\Delta_+=2$). By using Eq.~\eqref{eq:sol2}, the result for the static Green's function is
\begin{multline}
G(\omega=0)=\frac{1}{2}\Bigl[\psi\Bigl(\frac{1}{2}+\nu_k-\frac{q}{\sqrt{3}}\Bigr)\\
+\psi\Bigl(\frac{1}{2}+\nu_k+\frac{q}{\sqrt{3}}\Bigr)+2\gamma+\ln 3\Bigr],
\end{multline}
where $\psi(x)$ is the digamma function.

\textit{Special case 2}: $m^2=-3$ ($\Delta_+=3$). Similarly, the result for the static Green's function is
\begin{align}
G&(\omega=0)=(3\nu_k^2-q^2)\Bigl[\psi\Bigl(1+\nu_k-\frac{q}{\sqrt{3}}\Bigr)\nonumber\\
&+\psi\Bigl(1+\nu_k+\frac{q}{\sqrt{3}}\Bigr)+2\gamma+\ln 3-1\Bigr]+\frac{1}{2}-3\nu_k.
\end{align}

\textit{Special case 3}: $m^2=0$ ($\Delta_+=4$) and $q=0$. The solution for Eq.~\eqref{eq:KG} at $\omega=0$ is
\begin{equation}
\phi_O=C_1+C_2\left(\frac{3}{1-z^2}+\ln\frac{1-z^2}{2z^2+1}\right).
\end{equation}
The solution for Eq.~\eqref{eq:KG} in the inner region is $\phi_I=e^{i\omega/[12(1-z)]}$. We obtain $G(\omega=0)=0$, which implies $\kappa_c\to\infty$ at $u=3$, and thus is consistent with Fig.~\ref{fig:pdiag12}.

\textit{Special case 4}: $m^2=5$ ($\Delta_+=5$). This corresponds to $u=8$ in Fig.~\ref{fig:pdiag12}. The result is cumbersome, but it is clear that the Green's function at $\omega=0$ is a finite number, which implies that $\kappa_c\neq 0$. However, Fig.~\ref{fig:pdiag12} shows that $\kappa_c=0$ at $u=8$. This inconsistency can be explained as follows. If we take the limit that $\Delta_+$ approaches an integer, Eq.~\eqref{eq:sol} diverges, and we need a renormalization, which changes $\kappa_c$.

Let's take a more careful examination about the cases when $\Delta_+=2$, $3$, $\cdots$. When $\Delta_+=2$, i.e., $m^2=m_\text{BF}^2=-4$, $\phi=Az^2\ln z+Bz^2+\cdots$, where $A$ is the source. Now we consider $\Delta_+>2$. According to Eq.~\eqref{eq:a0b0}, the expectation value $B$ is divergent when $\Delta_+$ is an integer. If we add a small number $\delta$ to $\Delta_+$, we have
\begin{equation}
B=\frac{b}{\delta}+B_r+\mathcal{O}(\delta),
\end{equation}
where $B_r$ is the renormalized value of $B$. By the expansion
\begin{equation}
z^\delta=1+\delta\ln z+\frac{1}{2}(\delta\ln z)^2+\cdots,\qquad \delta\to 0,
\end{equation}
we have
\begin{align}
\phi=A &z^{\Delta_-}(1+\cdots)+Bz^{\Delta_++\delta}(1+\cdots)\nonumber\\
=A &z^{\Delta_-}(1+\cdots)\nonumber\\
+ &z^{\Delta_+}\Bigl(\frac{b}{\delta}+B_r+b\ln z+\mathcal{O}(\delta\ln z)\Bigr)(1+\cdots),
\end{align}
where ``$\cdots$" denotes higher-order terms in $z$. The renormalized Green's function by the standard quantization is $G_r=B_r/A$. Since there is a logarithm, the near boundary expansion should be evaluated at a UV cutoff $z=L_\text{UV}$ ($L_\text{UV}/L<<1$). Therefore, there are two noncommuting limits: $\delta\to 0$ and $L_\text{UV}\to 0$. The Green's function obtained by Eq.~\eqref{eq:a0b0} makes sense only if the following condition is satisfied:
\begin{equation}
\Bigl|\delta\cdot\ln\frac{L_\text{UV}}{L}\Bigr|<<1.
\end{equation}

\section{IR Green's function\label{sec:gir}}
The IR Green's function $\mathcal{G}_k(\omega)$ for the scalar field is given in Ref.~\cite{Faulkner:2009wj}. We will briefly review the result first. The IR geometry described by
\begin{equation}
f=12(1-z)^2,\qquad A_t=2\sqrt{6}(1-z).
\end{equation}
Define the AdS$_2$ coordinate
\begin{equation}
\zeta=\frac{1}{12(1-z)}.
\end{equation}
The solution to the Klein-Gordon equation with the in-falling boundary condition at the horizon is
\begin{equation}
\phi\sim W_{-\frac{iq}{\sqrt{6}},\nu_k}(-2i\omega\zeta),\label{eq:phiin2}
\end{equation}
where $W_{\lambda,\mu}(x)$ is a Whittaker function with the following asymptotic behavior:
\begin{equation}
W_{\lambda,\mu}(x)\sim e^{-x/2}x^\lambda(1+\cdots),\qquad |x|\to\infty.
\end{equation}
By expanding Eq.~\eqref{eq:phiin2} at $\omega\zeta\to 0$, we obtain
\begin{equation}
\phi=\zeta^{1/2-\nu_k}+\mathcal{G}_k(\omega)\zeta^{1/2+\nu_k}.
\end{equation}
The IR Green's function at zero temperature is
\begin{equation}
\mathcal{G}_k(\omega)=\frac{\Gamma(-2\nu_k)\Gamma(\frac{1}{2}+\nu_k-\frac{iq}{\sqrt{6}})}
{\Gamma(2\nu_k)\Gamma(\frac{1}{2}-\nu_k-\frac{iq}{\sqrt{6}})}(-2i\omega)^{2\nu_k}.\label{eq:gir}
\end{equation}
The finite temperature generalization at $T<<\mu$ (chemical potential) is
\begin{multline}
\mathcal{G}_k^{(T)}(\omega)=(4\pi T)^{2\nu_k}\\
\times\frac{\Gamma(-2\nu_k)\Gamma\bigl(\frac{1}{2}+\nu_k-\frac{iq}{\sqrt{6}}\bigr)
\Gamma\bigl(\frac{1}{2}+\nu_k-\frac{i\omega}{2\pi T}+\frac{iq}{\sqrt{6}}\bigr)}
{\Gamma(2\nu_k)\Gamma\bigl(\frac{1}{2}-\nu_k-\frac{iq}{\sqrt{6}}\bigr)
\Gamma\bigl(\frac{1}{2}+\nu_k-\frac{i\omega}{2\pi T}+\frac{iq}{\sqrt{6}}\bigr)}.\label{eq:girT}
\end{multline}
Near the bifurcating critical point $\nu\to 0$,
\begin{equation}
\mathcal{G}_{k=0}^{(T)}(\omega)=-1+2\nu\mathcal{G}_0^{(T)}(\omega),
\end{equation}
where
\begin{multline}
\mathcal{G}_0^{(T)}(\omega)=-\psi\Bigl(\frac{1}{2}+\frac{iq}{\sqrt{6}}-\frac{i\omega}{2\pi T}\Bigr)\\
-2\gamma-\ln(4\pi T)-\psi\Bigl(\frac{1}{2}-\frac{iq}{\sqrt{6}}\Bigr).\label{eq:gir0T}
\end{multline}
Particularly, for a neutral scalar,
\begin{equation}
\mathcal{G}_0^{(T)}(\omega)=-\psi\Bigl(\frac{1}{2}-\frac{i\omega}{2\pi T}\Bigr)
-\gamma-\ln(\pi T).\label{eq:gir00T}
\end{equation}

For $q>0$ and $\nu<1/2$, Eq.~\eqref{eq:C} can be proved by
\begin{align}
\arg\frac{1}{(-C)}=&\arg\left(-\frac{\Gamma(2\nu)\Gamma(\frac{1}{2}-\nu-iq_*)}
{\Gamma(-2\nu)\Gamma(\frac{1}{2}+\nu-iq_*)}\right)\nonumber\\
=&\arg[\cos\pi(\nu-iq_*)]\nonumber\\
=&\arctan(\tan\pi\nu\tanh\pi q_*)<\pi\nu,
\end{align}
where we have used the reflection formula $\Gamma(x)\Gamma(1-x)=\pi/\sin\pi x$, and the fact that we can put any positive number inside $\arg(x)$ without changing its value.

\end{document}